\documentstyle[12pt]{article}
\makeatletter
\@addtoreset{equation}{section}
\makeatother

\begin{document}
 \mbox{} \hspace{1.0cm}September 1992 \hspace{6.4cm}HLRZ-92-69\\
\begin{center}
\vspace*{1.0cm}
{{\large Chern-Simons term in the  \\
4-dimensional SU(2) Higgs Model}
 } \\
\vspace*{1.0cm}
{\large F.~Karsch$^{1,2}$,
        M.~L.~Laursen$^{1,}
$\footnote{Talk presented by M.~L.~Laursen at the Symposium on Lattice
Field Theory, Lattice 92, Amsterdam September 15-19, 1992.}  \\
        T.~Neuhaus$^2$ and B.~Plache$^2$} \\
\vspace*{1.0cm}
{\normalsize
$\mbox{}^1$ {HLRZ, c/o KFA J\"{u}lich,
             P.O. Box 1913, D-5170 J\"{u}lich, Germany}\\
$\mbox{}^2$ {Fak. f. Physik, Univ. Bielefeld,
              D-4800 Bielefeld, Germany}}\\
\vspace*{2cm}
{\large \bf Abstract}
\end{center}
\setlength{\baselineskip}{1.3\baselineskip}

Using Seibergs definition for the geometric charge in SU(2)
lattice gauge theory, we have managed to apply it also
to the Chern-Simons term.
We checked the periodic structure and
determined the Chern-Simons density on  small lattices $L^4$
and $L^3 \times 2,\, 4$
with $L=4,\,  6,\mbox{ and }8$
near the critical region in the SU(2) Higgs model. The data indicate
that tunneling is increased  at high temperature.
\newpage
\section{Introduction}

Some years back 't Hooft found that the baryon number and the
lepton number are not conserved in the electroweak theory \cite{Hooft}.
While the $B-L$ symmetry remains unbroken
due to the anomaly cancellation, $B+L$ is no
longer conserved. This socalled baryon number violation is
caused by the nontrivial topological winding of the SU(2) gauge fields.
The anomaly of the fermionic current relates the winding of the gauge
fields and changes  the baryon number by an amount
\begin{equation}
 B(t_2) - B(t_1) = \frac{N_f}{16\pi^{2}}
                   \int_{t_1}^{t_2} \int d^{3}x
                    tr[F_{\mu\nu}\tilde{F}_{\mu\nu}]
\end{equation}
where $N_f$ is the number of families of quarks and leptons.
In the axial gauge $A_0=0$
we can relate the change in the baryon number
to the change in the Chern-Simons number
\begin{equation}
            B(t_2)-B(t_1) = N_{f}[N_{CS}(t_2)-N_{CS}(t_1)]
\end{equation}
where the Chern-Simons number  $N_{CS}$ is
\begin{equation}
           N_{CS}= - \frac{1}{8\pi^{2}} \int d^{3}x
   \epsilon_{ijk} tr[A_{i}(\partial_{j}A_{k}+\frac{2}{3}A_{j}A_{k})].
\end{equation}
At zero temperature such processes are exponentially suppressed as
$exp(-2\pi/\alpha_W)$, $\alpha \approx 1/30$. This is because any
gauge field configuration which changes the winding number has an action
at least that of the barrier height
$2\pi/\alpha_W$.

At high temperatures which prevail in the early
universe such an exponential suppression is absent, since the system can
pass over the barrier classically. The only suppression factor is the
Boltzmann factor $exp(-\beta E)$ where $E$ is the barrier height,
and this factor is close to one ref.~\cite{Kuzmin}.
Any baryon asymmetry generated at the GUT scale will
get washed out as the universe approaches the electroweak phase
transition from above. If one assumes that the
transition is of first order, together with CP violating processes
and thermal non-equilibrium (provided by the expansion
of the universe) one can explain the baryon asymmetry in
the universe.

There exists semiclassical solutions of the
Yang-Mills Higgs fields which are believed to be important in this
scenario. These solutions are known as sphalerons. They are
static, have finite energy, but are unstable.
The sphaleron sits so to speak on top of the barrier and it has
a baryonic charge of 1/2.  While the instanton tunnels between two
Chern-Simons vacua  (in the axial gauge) one must imagine
a different time dependent solution which interpolates between
a Chern-Simons vacuum and the top of the barrier.

There are      some lattice studies of baryon number violating
processes in the 2d - Abelian Higgs model \cite{Grig},
and in the 4d - SU(2) Higgs model
\cite{Ambjorn}. The configurations
are prepared at high temperature and the system is allowed
to change via the classical Hamiltonian equation of motion.
Since the axial gauge is used, the Gauss
constraint must be implemented in addition.
The Chern-Simons term $N_{CS}(t)$  is monitored during the time evolution
as a function of the temperature. When
the system passes through a sphaleron transition one finds
$\Delta N_{CS} = \pm 1$. All these calculations are done in the
real time formalism.

We have initiated work on the 4d - SU(2) Higgs model in Euclidean
time, trying to see how the temperature influences tunneling.
To evaluate $N_{CS}$ we     have
used a geometric definition given by Seiberg see
ref.~\cite{Luesch}.
It requires a two dimensional numerical integration,
but we have an efficient vectorized code which
is an improvement of the code used in ref.~\cite{Fox}.
See also ref.~\cite{Gock} for alternative definitions of
the Chern-Simons term.

\section{Topological charge and the Chern-Simons term in the continuum}
We will first define the topological charge in SU(2).
The gauge field is $A_{\mu}$ and the gauge field tensor is
$F_{\mu\nu}  = \partial_{\mu}A_{\nu} - \partial_{\nu}A_{\mu}
      + [A_{\mu},A_{\nu}]$.
Under a a local gauge transformation $g$  the gauge field changes as:
$\delta A_{\mu} =  g^{-1}[A_{\mu} + \partial_{\mu}]g(x)$,
while the gauge field tensor transforms    gauge covariantly
$F_{\mu\nu} \rightarrow  g^{-1}F_{\mu\nu}g$.
The topological charge $Q$ is gauge invariant and an integer,
\begin{equation}
 Q = - \frac{1}{32\pi} \int_{M} d^{4}x
 \epsilon_{\mu\nu\rho\sigma}tr[F_{\mu\nu}F_{\rho\sigma}] \in Z.
\end{equation}
The manifold is denoted M and we shall assume that its boundary
$\partial M$ is a three sphere $S^3$.
The topological charge density $q$ can be written
as  a perfect derivative
\begin{equation}
 q = - \frac{1}{32\pi}
 \epsilon_{\mu\nu\rho\sigma}tr[F_{\mu\nu}F_{\rho\sigma}]
        =  \partial_{\mu}K_{\mu}
\end{equation}
where the Chern-Simons density $K_{\mu}$ is
\begin{equation}
 K_{\mu} = - \frac{1}{8\pi^{2}} \epsilon_{\mu\nu\rho\sigma}
   tr[A_{\nu}(\partial_{\rho}A_{\sigma}+
   \frac{2}{3}A_{\rho}A_{\sigma})].
\end{equation}
It is gauge variant and changes under the gauge transformation $g$
by an amount
(${\cal G}_{\nu} = \partial_{\nu}g\,g^{-1}$)
\begin{eqnarray}
 \delta K_{\mu} = & - & \frac{1}{24\pi^2}
  \epsilon_{\mu\nu\rho\sigma}
  tr[{\cal G}_{\nu}{\cal G}_{\rho}{\cal G}_{\sigma}] \nonumber \\
                                   & - & \frac{1}{8\pi^2}
  \epsilon_{\mu\nu\rho\sigma}\partial_{\nu}
  tr[{\cal G}_{\rho}A_{\sigma}].
\end{eqnarray}
We define the (timelike) Chern-Simons  number $N_{CS}$
as follows:
\begin{equation}
 N_{CS} = \int_{\partial M} d^{3}x K_{0} \not \in Z.
\end{equation}
While $N_{CS}$ is only an integer for pure gauge configurations,
the gauge variation is an integer (the boundary term vanishes)
\begin{equation}
 \delta N_{CS} = - \frac{1}{24\pi^2} \epsilon_{0\nu\rho\sigma}
  \int_{\partial M} d^{3}x
  tr[{\cal G}_{\nu}{\cal G}_{\rho}{\cal G}_{\sigma}]
 \in Z.
\end{equation}
This follows also from homotopy theory using the mapping
$g: S^{3} \rightarrow SU(2) = S^{3}$.
Such mappings are characterized with the homotopy class
$\Pi_{3}(S^{3}) \in Z$.

\section{Topological charge and the Chern-Simons term on the lattice}
We will now consider the lattice version  of the topological charge
and the  Chern-Simons number. We will use a geometric definition.
Problems with dislocations will be ignored here.
The manifold is a four torus
 $M = T^4$ and we will  cover $M$ with cells (hypercubes) $c(n)$.
Let the  gauge potential
$A_{\nu}^{n}$ be defined on $c(n)$ and likewise
$A_{\nu}^{n-\hat{\mu}}$ be defined on $c(n-\hat{\mu})$.
At the faces  (cubes)
$f(n,\mu) = c(n-\hat{\mu}) \cap c(n)$, we can relate
the two potentials by   a  transition function
$v_{n,\mu}$
\begin{equation}
  A_{\nu}^{n-\hat{\mu}}  =
  v^{-1}_{n,\mu} [A_{\nu}^{n} + \partial_{\nu}] v_{n,\mu}.
\end{equation}
In L\"{u}schers version one first fixes to a local complete axial gauge
in each $c(n)$. This will define  $v_{n,\mu}$ at the corners of
the hypercube. It is then possible to extend it to the whole
cube. The topological charge is
(${\cal S}_{\nu} =  s^{-1}\partial_{\nu}s,
  {\cal P}_{\nu} =  p^{-1}\partial_{\nu}p$),
\begin{eqnarray*}
Q^{L} = \sum_{n} q^{L}(n)   =
             \sum_{n,\mu}(-1)^{\mu}(k_{n,\mu} - k_{n+\mu,\mu}),
\end{eqnarray*}
\begin{eqnarray}
 (-1)^{\mu} k_{n,\mu} = & - &
 \frac{1}{24\pi^2} \epsilon_{\mu\nu\rho\sigma} \int_{f} d^{3}x
  tr[{\cal S}_{\nu}{\cal S}_{\rho}{\cal S}_{\sigma}] \nonumber \\
        &     + &
 \frac{1}{8\pi^2} \epsilon_{\mu\nu\rho\sigma} \int_{\partial f} d^{2}x
  tr[{\cal P}_{\rho}{\cal S}_{\sigma}].
\end{eqnarray}
The function $s$ is defined on the cube,
while        $p$ is defined on the boundary of the cube. The actual
expressions are given in ref.~\cite{Luesch}.
In Seibergs version no local gauge fixing is performed, but otherwise
the same interpolation is performed. Replace
$(s,p,k_{n,\mu}) \rightarrow  (S,P,K_{n,\mu})$. The difference is that
that $S$ and $P$ only depend on the original gauge fields in the cube.
Then
$N_{CS} = \sum_{n_s} K_{n_{s},\mu}$
is nothing but a Chern-Simons term
(the summation is over the spatial lattice only).
Like in the continuum
it is only an integer for pure gauge field configurations, but under
gauge transformations it changes by an integer. The
corresponding topological charge is defined as
$Q^{S}   =  \sum_{n} \tilde q^{S}(n)$,
where $-1/2 \leq \tilde q^{S}(n) <    1/2$.
By restricting the charge to this interval we will have a gauge invariant
charge definition.
We find that both the topological charge and the Chern-Simons term
have the correct naive continuum limit.
As an interesting corollary we verify that the two charge definitions
are related. If we introduce the section of the L\"{u}scher bundle:
$w(x)$, $x \in \partial c(n)$,
then $s(x) = w(0)S(x)w^{-1}(x)$
Inserting this in
the above expressions yields $q^{S}(n) = q^{L}(n) - q^{w}(n)$
where $q^{w}(n)$ is the topological charge (integer) of the section.
Therefore $\tilde q^{S}(n) = q^{L}(n)$ up to integers
\cite{Lau}.
For smooth fields like instantons they always agree, while for realistic
configurations this is true for almost every hypercube.
It is interesting to look at the
Chern-Simons number at each time slice of an instanton in the complete
axial gauge. This is shown in Fig.~\ref{fig:instanton}.
We start in the vacuum sector $N_{CS} = 0$ and move towards
the other vacuum sector $N_{CS} = 1$.
On the last time slice we must return to the
first vacuum due to periodic boundary conditions.

\section{Tests and Monte-Carlo results for the Chern-Simons density}

We have used an SU(2) Higgs model with action:
\begin{eqnarray}
  S = & - & \frac{\beta}{2} \sum_{n,\mu <\nu}tr[U_{n,\mu\nu}]
        +   \lambda \sum_{n}(\Phi^{\dagger}_{n}\Phi_{n} - 1)^2
      \nonumber \\
    & - & \kappa \sum_{n,\mu}tr[\Phi^{\dagger}_{n}U_{n,\mu}\Phi_{n+\mu}].
\end{eqnarray}
We always tried to work close to the Higgs phase transition.
We looked at the periodic structure for a $4^3\times 2$ lattice at
$(\beta,\kappa,\lambda) = (2.25,0.3,0.5)$.
We did 1000 configurations and we measured $N_{CS}$ without any
gauge fixing.
In Fig.~\ref{fig:periodic}
we have plotted the Chern-Simons probability. The periodic
structure is obvious. Notice that most of the configurations
have $N_{CS}$ close to an integer, and therefore can be
interpreted as being  pure gauge.
We checked that $N_{CS}$ indeed changed  by an integer under an axial
gauge transformation.
Next we compared
the probabilities for the two lattices $6^4$ and $6^{3}\times 2$.
For these lattices we used 50 Landau gauge fixing
sweeps, to make the integrals converge fast.
In both cases we have 6000 Chern-Simons numbers.
The results for the restricted and therefore gauge invariant
$\tilde N_{CS}$
($ -1/2 \leq \tilde N_{CS} < 1/2$) are shown in
Fig.~\ref{fig:zerot}
and Fig.~\ref{fig:finitet}.
There is a trend in the direction of a
flatter distribution at finite temperature.
We take this as evidence that the system tunnels more often.
A few words about the integrations done.
One of the integrations can be done analytically
so we are left with a two dimensional integral.
We have used  the following strategy, which turned out to be quite
efficient. Perform a Gaussian integration with $8\times 8$ points and
store the results for the eight $K_{n,\mu}$'s in each hypercube.
Redo the same thing with $16\times 16$ points and compare the results
for each $K_{n,\mu}$. If the relative
difference is less than 0.0001 we accept the contribution.
Otherwise we collect the
$K_{n,\mu}$'s which
have not yet converged. These we redo with
$32\times 32$ points instead.
Compare with the previous values and repeat the procedure with
$64\times 64$ points. Usually, at this point only a few
$K_{n,\mu}$'s have not converged,
so for these we use a library integration
routine with interval adaption. The typical time for one topological
charge on a $6^4$ lattice is 100 seconds on the CRAY-YMP.
The charges are integers up to errors of the order $10^{-4}$.
The corresponding time for one Chern-Simons number in a timeslice
is around 5 seconds.

\begin{figure}[t]
\caption{Profile of
$N_{CS}$ through an instanton configuration.}
\label{fig:instanton}
\end{figure}
\begin{figure}[b]
\caption{The periodic structure for $N_{CS}$.}
\label{fig:periodic}
\end{figure}
\begin{figure}[t]
\caption{
The Chern-Simons probability  at zero temperature
as a function of $\tilde N_{CS}$.}
\label{fig:zerot}
\end{figure}
\begin{figure}[b]
\caption{
The Chern-Simons probability  at finite temperature
as a function of $\tilde N_{CS}$.}
\label{fig:finitet}
\end{figure}

\newpage

\begin{thebibliography}{99}
\bibitem{Hooft} G.~'t Hooft, Phys.~Rev.~Lett. 37 (1976) 8;
                   Phys.~Rev.~{\bf D14} (1976) 3432.
\bibitem{Kuzmin} V.~A.~Kuzmin, V.~A.~Rubakov and M.~E.~Shaposhnikov,
                 Phys. Lett. {\bf B155} (1985) 36.
\bibitem{Grig} D.~Yu.~Grigoriev, V.~A.~Rubakov and M.~E.~Shaposhnikov,
               Phys.~Lett.~{\bf B216} (1989) 172;
               Nucl.~Phys.~{\bf B326} (1989) 737; \\
                  A.~I.~Bochkarev and Ph.~de~Forcrand,
                  Phys.~Rev.~{\bf D44} (1991) 519.
\bibitem{Ambjorn} J.~Ambj\o rn, M.~L.~Laursen and M.~E.~Shaposhnikov,
                  Phys.~Lett.~{\bf B179} (1987) 757;
                  Nucl.~Phys.~{\bf B353} (1989) 483;
                  J.~Ambj\o rn {\it et al},
                  Phys.~Lett.~{\bf B244} (1990) 479;
                  J.~Ambj\o rn, K.~Farakos,
                  NBI-HE-92-52 preprint.
\bibitem{Luesch}M.~L\"{u}scher, Commun.~Math.~Phys.{\bf 85} (1982) 39;
                 N.~Seiberg, Phys.~Lett.~{\bf B148} (1984) 456.
\bibitem{Fox} I.~A.~Fox, J.~P.~Gilchrist, M.~L.~Laursen and G.~Schierholz,
              Phys.~Rev.~Lett.{\bf 54} (1985) 749.
\bibitem{Gock} M.~G\"{o}ckeler, A.~S.~Kronfeld, G.~Schierholz and
              U.~-J.~Wiese, HLRZ-92-34 preprint; \\
              A.~Phillips and D.~Stone,
               Nucl.~Phys.~{\bf B} (Proc. Suppl.) 20 (1991) 28.
\bibitem{Lau} M.~L.~Laursen, HLRZ-92-61 preprint.
\end{thebibliography}
\end{document}